\documentclass[sigconf]{acmart}

\usepackage{tcolorbox}
\tcbuselibrary{skins}
\usepackage{multirow}
\usepackage{booktabs}
\usepackage{tabularx}
\usepackage{bbding}
\usepackage{graphicx}
\usepackage{caption}
\usepackage{subcaption}
\usepackage{xcolor}
\usepackage{textcomp}
\usepackage{algorithmic}
\usepackage{algorithm}
\usepackage{xurl}

\newtcolorbox{promptbox}[1]{
  colback=gray!5,      
  colframe=black,      
  fonttitle=\bfseries, 
  title={#1},          
  sharp corners,       
  boxrule=0.5pt        
}

\AtBeginDocument{%
  }

\setcopyright{acmlicensed}
\copyrightyear{2026}
\acmYear{2026}
\acmDOI{XXXXXXX.XXXXXXX}
\acmConference[Conference acronym 'XX]{Make sure to enter the correct
  conference title from your rights confirmation email}{June 03--05,
  20}{Woodstock, NY}
\acmISBN{978-1-4503-XXXX-X/2018/06}

\begin{document}




\title{TextClusterLab: An Integrated Framework for Reliable Text Clustering Studies}

\author{Daoming Wan, Yizheng Huang, and Jimmy X. Huang}
\authornote{Corresponding author.}
\affiliation{%
  \institution{Information Retrieval and Knowledge Management Research Lab, York University}
  \city{Toronto}
  \state{Ontario}
  \country{Canada}
}
\email{{daoming, hyz, jhuang}@yorku.ca}







\renewcommand{\shortauthors}{Wan et al.}

\begin{abstract}
In recent years, text clustering has become a critical technique for applications including intent discovery, topic mining, and recommendation systems. However, evaluating text clustering algorithms remains challenging since many real-world textual datasets are not suitable for clustering assessment due to ambiguous semantic boundaries, the high dimensionality of embeddings, and inconsistent cluster structure. Current clustering dataset generators are designed for numerical data, providing limited support for text-specific benchmarking. This paper introduces TextClusterLab, a comprehensive framework for text clustering research. TextClusterLab offers a Large Language Model (LLM)-driven text clustering dataset generator to produce synthetic text datasets for evaluating clustering algorithms. This generator supports setting various clustering attributes, such as class imbalance, intra-cluster compactness, and inter-cluster diversity. These generated datasets can serve as practical benchmarks for testing the robustness and versatility of text clustering algorithms in diverse scenarios. Moreover, we introduce a benchmark to verify whether a text dataset is suitable for clustering evaluation. Therefore, TextClusterLab provides an integrated framework for reproducible and comprehensive text-specific clustering research. Our TextClusterLab is publicly available on \href{https://github.com/research-paper-code/TextClusterLab}{GitHub}, and some synthetic example datasets with various attributes are publicly available on \href{https://huggingface.co/datasets/DW-irlab/TextClusterLab}{Hugging Face}.
\end{abstract}



\received{20 February 2007}
\received[revised]{12 March 2009}
\received[accepted]{5 June 2009}

\maketitle

\section{Introduction and Motivation}



In recent years, the exponential growth of unsupervised textual content has made text clustering a fundamental component of many natural language processing (NLP) applications \cite{park2022toward}. By revealing hidden structure in unlabeled textual data, text clustering groups a corpus based on semantic similarity. Effective partitions of textual datasets are beneficial for a range of downstream tasks, including intent discovery, topic mining in social media, and recommendation systems \cite{aggarwal2012survey, ahmed2022short, hong-etal-2025-dial}. In this case, the development and selection of high-performance clustering algorithms play crucial roles in obtaining high-quality clustering results.


However, in benchmarking clustering algorithms, the choice of test data has a significant impact on the measured performance \cite{jain1999data}. When the dataset has no inherent clustering structure, no clustering algorithm can be expected to produce reliable or stable results \cite{adolfsson2019cluster}. Therefore, selecting appropriate experimental datasets is a critical component in designing clustering studies. The comparative experiments on inappropriate datasets can lead to necessarily arbitrary results from any clustering algorithm \cite{ben2006sober}, reducing the effectiveness and reliability of the evaluation. On the other hand, a dataset containing a large amount of noisy samples can distort the accuracy of clustering results \cite{goyal2024systematic, 10.1145/3689036}.

In practice, most real-life domain-specific textual datasets inherently contain issues that affect their clusterability, such as ambiguous or overlapping semantic categories, heavy label noise, context-poor utterances, class imbalance, and multi-intent or multi-topic samples \cite{park2025dynamic, casanueva2022nlu++, wu2025multi}. These issues prevent the textual dataset from learning a clear structure in its latent representations, limiting effective semantic-level segmentation \cite{ortiz2023does}. In this case, evaluating whether a dataset is suitable for cluster analysis and synthesizing proper clusterable datasets from real-world data have become critical challenges for enabling reliable clustering studies \cite{ren2024deep}.

Existing methods for synthesizing clusterable datasets mainly focus on numerical datasets \cite{aziira2020generation}. Compared to textual data, numerical features are feasible to estimate data distributions (e.g., Gaussian mixtures) and easily control the construction of new samples that have preset clustering properties in terms of compactness, separation, and imbalance \cite{melnykov2012mixsim, shand2021hawks}. In contrast, text clustering is typically conducted on latent representations generated by embedding models \cite{chen2014mining}, which are often high-dimensional. The curse of dimensionality makes it difficult to interpret geometric shapes, observe a reliable density, and estimate a feasible distribution that can construct new samples \cite{yin2014dirichlet}. Moreover, even if the textual dataset has a clear structure in the latent space, embeddings produced by encoder-only models are inherently non-invertible. Semantic information may be compressed or lost in the encoding procedure, and mapping latent representations back to valid, faithful texts is non-trivial \cite{li2023sentence}. While some encoder–decoder models provide a pipeline for textual data generation, controlling fine-grained clustering properties through the operation of latent representations is still a challenge, and it is difficult to ensure the feasibility and semantic faithfulness when synthetic text data is driven solely by latent representations \cite{long-etal-2024-llms, petukhova2025text}.

To address these challenges, this paper proposes an integrated framework for reliable and comprehensive text-specific clustering studies. First, we introduce a benchmark to thoroughly evaluate the clusterability of a textual dataset, enabling researchers to identify their potential cluster properties prior to evaluating clustering algorithms. Second, we propose an LLM-based dataset generator that synthesizes text datasets designed for clustering, supports controllable clustering attributes (e.g., class imbalance, intra-cluster compactness, and inter-cluster diversity), and preserves the ground-truth semantic information of real-life data. In our generator, high-dimensional embeddings are used only for example selection, enabling the construction of high-relevant prompts for the LLM.  Finally, we design an evaluation protocol that benchmarks the robustness and versatility of clustering algorithms under various clustering attributes of domain-specific text datasets. These components provide a practical foundation for reproducible and comprehensive evaluation of text clustering methods under complicated and dynamic scenarios. The key contributions of this paper are listed as follows:
\begin{itemize}
  \item A novel benchmark is proposed to evaluate the clusterability of text datasets, helping to avoid unreliable experimental conclusions in algorithm comparisons.
  \item An LLM-driven generator produces clustering-friendly text datasets from real-life data, with easy control over clustering attributes such as class imbalance, intra-cluster compactness, and inter-cluster diversity.
  \item The standardized evaluation protocol is designed to comprehensively test clustering methods under diverse clustering attributes, enabling reasonable comparison of robustness and versatility.
  \item We integrate clusterability evaluation, dataset synthesis, and algorithm benchmarking into a complete framework, making text clustering research more systematic, practical, and reproducible.
\end{itemize}

\section{Related Work}

Existing text clustering methods typically follow an encode-cluster pipeline. The embedding or vectorizer method (e.g., Bag-of-Words, one-hot, and TF-IDF) is applied to map the text datasets into latent representations, and then the clustering algorithm segments latent representations and assigns the corresponding text samples into clusters \cite{aggarwal2012survey}. While such encoders can achieve reasonable clustering quality, the compression steps may neglect inherent semantic information that is essential for producing fine-grained, high-quality partitions. To address this limitation, recent studies have explored LLM-derived embeddings for clustering. Petukhova et al. \cite{petukhova2025text} systematically compares classical baselines (TF-IDF and BERT \cite{devlin-etal-2019-bert}) against multiple LLM embedding methods (including closed and open-source LLMs) in several clustering algorithms and datasets. Their experimental results indicate that LLM embeddings usually have better performance to capture structured semantic information, improve cluster purity, and make silhouette scores more informative. Moreover, this research proves that solely increasing embedding dimensionality or simply applying summary-based dimensionality reduction does not consistently improve clustering quality.

Another application of LLMs to text clustering is using LLMs to refine the clustering results after an initial clustering step, especially to judge ambiguous assignments near cluster boundaries. CLUSTERLLM \cite{zhang-etal-2023-clusterllm} follows this idea by querying an instruction-tuned LLM with clustering perspective. It performs a triplet task (one anchor and two candidates) to decide which candidate is more semantically consistent with the anchor under the primary clustering results, and then the predicted triplets are used to fine-tune small embedders. Moreover, it designs a pairwise question to judge if two samples belong to the same category, which helps to determine an appropriate clustering granularity. Another similar work, LLMEdgeRefine \cite{feng-etal-2024-llmedgerefine}, focuses on boundary/outlier repair. After attaining initial clustering, it identifies edge points and forms super-points from boundary samples to mitigate the influence of outliers and performs secondary re-clustering. It then applies LLM-Assisted Cluster Refinement (LACR) to identify some farthest edge points and removes or reassigns them based on their semantic context. It shows its advantage in using fewer prompts to improve semantic coherence, compared to other heavy querying baselines.

Recently, LLMs have been used to generate synthetic datasets for cluster analysis, realized by mapping human verbal descriptions into simulated clustering parameters of synthetic datasets. In particular, the repliclust framework \cite{zellinger2025natural} generates synthetic benchmarks directly from high-level user-defined archetypes (e.g., desired overlap, radius diversity, class imbalance, and distribution proportions). It summarizes the clustering parameter of probabilistic mixture-model geometry, and uses few-shot prompts to map the verbal description into parameter settings. Additionally, it supports post-processing to create irregular geometric structures for various scenarios. While this approach enables controllable benchmark generation for clustering analysis, it is currently limited to numerical features, and effectively synthesizing textual datasets with controllable attributes remains an open challenge.

The latest clustering methods apply LLMs to synthesize text for clustering purpose. The generated text serves as additional information to represent the inherent knowledge in documents before applying conventional embedding-based clustering. Du et al. propose Information-Theoretic Generative Clustering (GC) \cite{du2025information}, which groups a document set by the additional text generated from LLMs, and then defining document–cluster similarity by KL divergence. This approach reports significant improvements over other clustering baselines, and it shows that this method can be adapted to generative document retrieval, where documents are indexed through hierarchical clustering. However, this method has mainly been evaluated on a dataset with few potential clusters. As the number of clusters grows, estimating the required probability matrix becomes exponentially expensive. Moreover, the generated text (produced by the doc2query model \cite{nogueira2019document}) can only be treated as additional information of a certain document. Its dataset-level clustering attributes are fixed and unchangeable.

\section{Data Clusterable Benchmark}

This section explains the principle to build a Data Clusterable Benchmark, introduces some criteria to construct the benchmark, and provides some examples to demonstrate it.

\subsection{Clusterable Benchmark Principle}

Appropriate experimental text datasets play a vital role in text clustering research. In practice, many widely used datasets have various limitations that present an unstructured distribution in the latent space. Therefore, it is necessary to evaluate the clusterability of the dataset before conducting experiments intended to test the performance of clustering algorithms. 

To build a reliable benchmark, we design criteria that satisfy three key requirements: (1) high-dimensional robustness, since experiments are typically performed in embedding spaces with 1,000+ dimensions, (2) stability, the scores of datasets remain consistent under resampling, and (3) structure identification, the ability to detect meaningful graph or community structure in the dataset. According to these requirements, our benchmark includes criteria for eight aspects: reliability, structure, stability, separation, agreement, coherence, imbalance, and meaningfulness.

\subsection{Benchmark Introduction}

For a given text dataset $\mathcal{D} = \{ x_i \}_{i=1}^{n}$, each sample is mapped into a latent representation by a strong encoder model (${e}_i = f(x_i)$). The benchmark produces a set of criteria results and normalizes them to $Score \in [0, 100]$. A radar chart is also provided to intuitively visualize the criteria scores, offering a basic summary of its properties. The details of each criterion are introduced as follows:

\textbf{Reliability} (C1): This criterion is used to check if the neighbors of a sample are meaningfully closer than random samples. For each sample $i$, $Sim(knn_i)$ stands for the average cosine distance between the given sample and its neighbors ($cos(e_i,e_j)$), and $Sim(rand_i)$ stands for the average cosine distance between the given sample and some random samples. 
The difference between these two similarities shows whether the dataset has reliable neighborhoods. The C1 score is computed as the sigmoid of the effect size:
\begin{equation}
    C1 = sigmoid((Sim(knn_i)-Sim(rand_i))/std(rand_i))
    \label{eq1}
\end{equation}
The higher score demonstrates that the datasets have strong local neighborhoods.

\textbf{Structure} (C2):  This criterion checks if the datasets have community structure in a neighbor graph. The KNN graph is created by edges, the $edge(i,j)$ exists only if sample $i$ and sample $j$ are in the mutual neighborhoods. The community detection algorithm \cite{yang2016comparative} is applied to compute the modularity of the partition and average conductance per community. The C2 score is the average of modularity and $(1-conductance)$. The higher score proves the dataset has an obvious community structure.

\textbf{Stability} (C3): This criterion evaluates if the cluster results are stable under bootstrap sampling. Two clustering methods (e.g., k-means and HDBSCAN) are selected to group the dataset for bootstrap $B$ times. In each time, randomly extract 80\% samples, cluster them by the same hyperparameters, and record the Adjusted Rand Index (ARI). The co-assignment is used for computing how often a pair of samples is assigned to the same cluster. The C3 score is computed as the average of the ARI from two clustering methods and co-assignment.
\begin{equation}
\small
C3 = mean(mean(ARI_{1}), mean(ARI_{2}), mean(coassignment)))
\label{eq2}    
\end{equation}

The higher score demonstrates that the datasets are stable for different clustering methods under resampling.

\textbf{Separation} (C4): This criterion detects if the produced clustering results have meaningful intra-cluster compactness structure and inter-cluster separation. The intra-cluster compactness is the mean value of the cosine distance between samples and their cluster centroid.
\begin{equation}
intra = \frac{1}{N}\sum_{c=1}^{C}\sum_{i\in S_c}\left(1-\cos\left(e_i,\mu_c\right)\right)
    \label{eq3}
\end{equation}
where $S_c$ and $\mu_c$ stand for the sample set, and the centroid of cluster $c$, respectively. $N$ is the sample size of the cluster.
The inter-cluster separation is the mean cosine distance between different cluster centroids.
\begin{equation}
inter=\frac{2}{N_C(N_C-1)}\sum_{c\in C}\sum_{d \in C, d \neq c} \cos\!\left(\mu_c,\mu_d\right)
    \label{eq4}
\end{equation}
where $N_C$ is the number of clusters. The C4 score is computed as the sigmoid of the logarithm of the ratio between intra-cluster and inter-cluster measures.
\begin{equation}
    C4 = sigmoid(log(inter / intra))
    \label{eq5}
\end{equation}
A higher score indicates that clusters are more compact and have clearer separation from other clusters.

\textbf{Agreement} (C5): This criterion checks the similarity of cluster structure in two different embeddings. For each sample, the kNN Jaccard index (intersection over union) presents the overlap rate of the kNN graph between two embeddings.
\begin{equation}
J_i = \frac{\left|kNN_A(i)\cap kNN_B(i)\right|}{\left|kNN_A(i)\cup kNN_B(i)\right|}
    \label{eq6}
\end{equation}
where $kNN_A(i)$ and $kNN_B(i)$ stand for the kNN results of embedding method A and B, respectively.
Then, $ARI \left(C^{A},\, C^{B}\right)$ is calculated as the ARI between clusterings from two embeddings. The C5 score is the average of kNN Jaccard and ARI between clusterings. The higher score presents the better structure agreement between different embeddings.

\textbf{Coherence} (C6): This criterion evaluates if the inside cluster samples always maintain similarity and coherence. This criterion is calculated by the average value of the silhouette score (SI) and the within-cluster similarity. The silhouette score presents the density of each cluster,
\begin{equation}
    s_i = \frac{b_i - a_i}{\max(a_i,\, b_i)}
    \label{eq7}
\end{equation}
and within-cluster similarity checks coherence inside the clusters.
\begin{equation}
\mathrm{wcs}_c = \frac{1}{N}\sum_{(i,j)} \cos\!\left(e_i, e_j\right)
    \label{eq8}
\end{equation}
where $e_i$ and $e_j$ stand for two different embedding samples. The higher scores indicate greater cluster density and higher internal similarity.

\textbf{Imbalance} (C7): This criterion quantifies dataset imbalance. For supervised datasets, it is defined as the ratio between the sample size of the largest class label and the sample size of the smallest class label. For unsupervised datasets, after an initial clustering step, it is defined as the ratio between the sample size of the largest cluster and the sample size of the smallest cluster. The C7 score is the inverse of the imbalance ratio (max/min). The higher score indicates that the sample sizes of the clusters/labels are closer to each other.

\textbf{Tendency} (C8): This criterion verifies whether the dataset has a real cluster structure, or if it randomly gets partitions in the high-dimensional latent space. The Hopkins statistic is used to calculate this criterion. It randomly draws $m$ samples from the datasets $s = \{ s_i \}_{i=1}^{m}$, and generates $m$ uniform random samples $R = \{ r_i \}_{i=1}^{m}$ within the same latent space bound. For each $s_i$, compute the distance to its nearest samples.
\begin{equation}
w_i = \min_{x \in X \setminus \{s_i\}} cos(s_i, x)
    \label{eq9}
\end{equation}
For each $r_i$, compute the distance to its nearest samples in the dataset.
\begin{equation}
u_i = \min_{x \in X} cos(r_i, x)
    \label{eq10}
\end{equation}
The Hopkins statistic is presented as follows:
\begin{equation}
H = \frac{\sum_{i=1}^{m} u_i}{\sum_{i=1}^{m} u_i + \sum_{i=1}^{m} w_i}
    \label{eq11}
\end{equation}
The C8 score is defined as $H \times 100$, and values close to 100 indicate a strong tendency toward clustering structure.

\subsection{Example Results}
The proposed benchmark is applied to four intent-based textual datasets. The details of these four datasets are presented in Appendix D.1. The scores are computed using Instructor embeddings and $k$-means clustering, with C3 additionally evaluated using agglomerative clustering. Scores of these four datasets are shown in Table \ref{table 1}. It clearly demonstrates that CLINC and Banking77 are more clusterable.

\begin{table}[h]
\centering
\caption{Clusterability criteria scores (C1--C8) for four datasets (CLINC, Banking77, Massive, and Mtop).}
\label{tab:clusterability_scores}
\begin{tabular}{ccccc}
\toprule
 & CLINC & Banking77 & Massive & Mtop \\
\midrule
Reliability (C1) & 99 & 99 & 98 & 99 \\
Structure (C2)   & 92 & 86 & 87 & 90 \\
Stability (C3)   & 89 & 87 & 87 & 92 \\
Separation (C4)  & 57 & 61 & 52 & 48 \\
Agreement (C5)   & 61 & 58 & 57 & 68 \\
Coherence (C6)   & 72 & 74 & 72 & 72 \\
Imbalance (C7)   & 100 & 100 & 0 & 0 \\
Tendency (C8)    & 81 & 80 & 75 & 74 \\
\bottomrule
\label{table 1}
\end{tabular}
\end{table}

The first three criteria (Reliability, Structure, and Stability) and the sixth criterion (Coherence) consistently obtain high values for all datasets. This result indicates that the embedding spaces of the four datasets are stable for different clustering methods and present a clear, well-defined cluster structure.

The fourth criterion, Separation, is weak for all datasets, indicating overlap between samples of different clusters in the high-dimensional latent space. As a result, it is difficult to obtain fine-grained and well-separated clustering results on these datasets.

The fifth criterion, Agreement, looks decent for all datasets, indicating that similar clustering structures for different embedding methods. This suggests that the clustering results are comparatively insensitive to the specific choice of encoder.

The Imbalance criterion presents only the variation in the sample size of the label (or pseudo-label). CLINC and Banking are class-balanced, with the same sample size for labels. In contrast, Massive and Mtop are highly imbalanced. For Massive, the largest cluster contains 209 samples, while the smallest contains 1. For MTOP, the largest cluster contains 473 samples, and the smallest contains 1.

The Tendency criterion is higher for CLINC and Banking77 than for Massive and Mtop, suggesting that CLINC and Banking77 present a more definite and meaningful cluster structure.

It is worth noting that not all criteria have a strong correlation with clusterability. For example, an imbalanced dataset may still present a well-separated cluster structure, whereas a label-balanced dataset can contain massive overlapping samples, especially pseudo-labels from unsupervised datasets. Nevertheless, the introduced benchmark provides a useful reference for the generator to synthesize more clusterable datasets by applying targeted strategies to improve certain criteria scores.

\begin{figure}[htbp]

  \centering
  \includegraphics[width=0.3\textwidth]{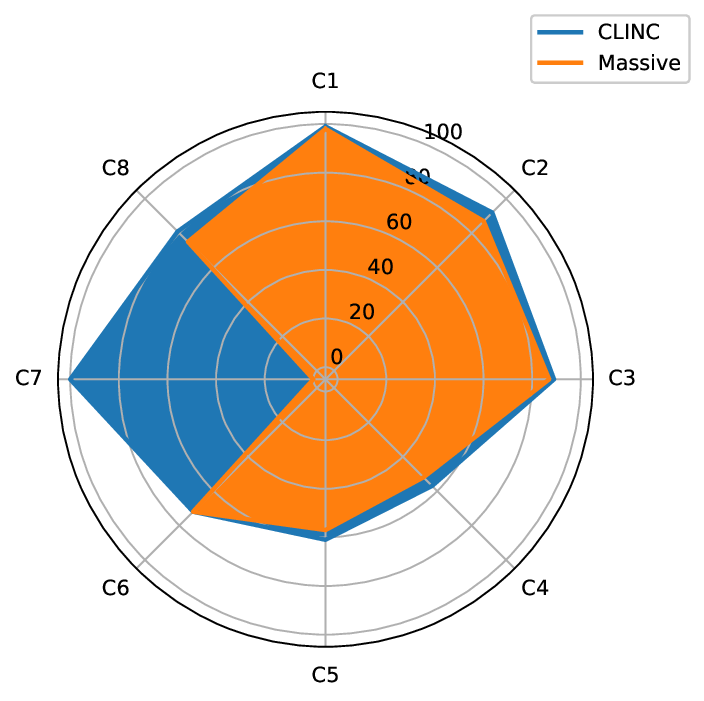}
    \caption{Benchmark radar chart of CLINC and Massive}
    \label{fig 1}

\end{figure}

The radar chart compares CLINC and Massive for eight criteria C1-C8 in a single view, presented in Figure \ref{fig 1}. The larger polygon generally means stronger overall performance. It obviously shows that the CLINC dataset has higher clusterability.

\section{Text Clustering Dataset Generation}

\begin{figure*}[!htbp] 
\centerline{\includegraphics[width=0.8\textwidth]{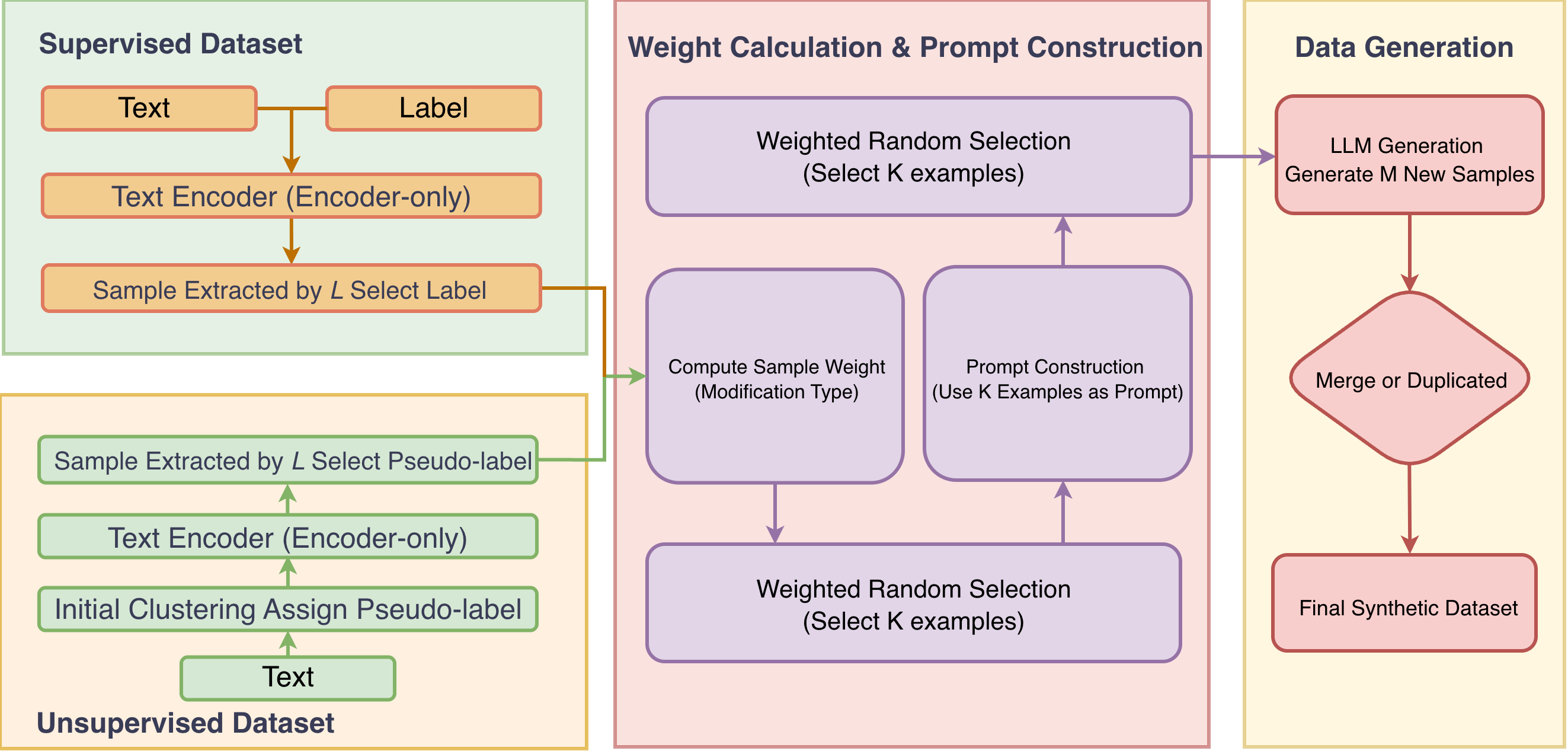}}
\caption{Supervised and unsupervised pipelines for LLM-driven text dataset synthesis. For supervised data, the textual data includes actual labels; for unsupervised data, pseudo-labels are assigned via an initial clustering step. The whole datasets are mapped to latent representations by an encode-only text encoder. The samples are extracted based on a preset number of labels/pseudo-labels. Computing weights for all extracted samples, and weighted randomly selecting $K$ samples of each label as examples to construct the prompt. The LLMs are used to generate textual datasets to build the final synthetic datasets.}
\label{fig 2}

\end{figure*}

This section introduces the pipeline of the text clustering dataset generator. The main components of this generator include text encoding, sample extraction, weight assignment, weighted random selection, prompt construction, and text generation. The workflow of this generator is shown in Figure \ref{fig 2}, and the examples of synthetic datasets are listed in Table \ref{table 8}.

\subsection{Development Procedure}

For a given text dataset $\mathcal{D} = \{ x_i \}_{i=1}^{n}$, if ground-truth labels are available, denote the labeled set as $\mathcal{D} = \{(x_i, y_i)\}_{i=1}^{N}$. Otherwise, an initial clustering algorithm is applied to obtain pseudo-labels and form $\mathcal{D} = \{(x_i, \tilde{y}_i)\}_{i=1}^{N}$. After that, each sample is mapped into a latent representation by using an encode-only model ${e}_i = f(x_i)$. The latent representations provide an initial clustering view of the dataset and serve as the basis for further example selection. 

Given a user-defined setting, the user specifies the number of labels ($N_L$) 
to be used for generating new samples. If the user provides a selected label list $ \tilde Y = \{y_i\}_i^{N_L}$, the whole samples that labels belong to the selected label list are extracted $S = \{(x_i, y_i)\mid y_i \in \tilde{Y}\}$. Otherwise, randomly select $N_L$ labels to create the label list. 

For the user-defined synthesis type setting, users can control the direction of synthetic data generation. We consider three synthesis types (imbalance, compactness, and diversity) and compute the sampling weight for each sample based on the selected synthesis type. Let $y_i \in \{1,...,C\}$ be the total label list. For each label $c$, the centroid of each class is computed as:
\begin{equation}
\mu_c=\frac{1}{|S_c|}\sum_{j\in S_c} e_j,\quad \text{where } S_c=\{x_j:\, y_j=c\}
    \label{eq12}
\end{equation}

\textbf{Imbalance} To control label-level imbalance, the boundaries of each cluster should be consistent. Therefore, the samples that have the same label or pseudo-label are assigned the same weight: $w_i = 1$. Within each label, every sample has the same chance to be selected as an example to construct prompts.

\textbf{Compact} To encourage a more compact cluster structure, a higher weight is assigned to the sample that is closer to its class centroid. The weight is defined by the Euclidean distance to the centroid:
\begin{equation}
w_i=\frac{1}{\left\lVert e_i-\mu_c\right\rVert_2+\epsilon},y_i = c
    \label{eq13}
\end{equation}
where $\epsilon >0$ is set to avoid division by zero. 

\textbf{Diversity} To promote cross-class informative generation and a looser clustering structure, higher weights are assigned to the samples closer to the cluster boundaries. The weight of a sample is defined as its total distance to the centroids of other labels:
\begin{equation}
w_i=\sum_{\substack{c=1\\ c\neq y_i}}^{C}\left\lVert z_i-\mu_c\right\rVert_2
    \label{eq14}
\end{equation}
This weighting increases the probability that boundary samples are selected as examples for prompt construction.

In practice, the weights $w_i$ are converted into a sampling distribution for weighted random selection of prompt examples. The equations of the normalized sampling distribution for sample $i$ are shown as follows:
\begin{equation}
\gamma=\frac{\log N_w}{\log\!\left(\frac{w_{\max}}{w_{\min}}\right)},\qquad
p_i=\frac{w_i^{\gamma}}{\sum_{j=1}^{N_w} w_j^{\gamma}}
    \label{eq15}
\end{equation}
where $N_w$ is the size of the weights, and $w_{\max}$, $w_{\min}$ are max/min values of weights, respectively. 

For each user-defined label, the weighted random selection is performed to choose $K$ prompt examples from its sample set. Specifically, we first obtain a sharpened version of the normalized sampling distribution $p_i$, and then $K$ examples are drawn without replacement according to the sampling distribution $p_i$. Samples with larger $p_i$ are more likely to be selected while still preserving randomness. The selected examples are then used to construct the LLM prompt. The example prompt is listed in Appendix A.

The large language model generates a user-specified number $M$ of new samples with a constructed prompt. The generated samples can either be merged with the original datasets or duplicated (replace original samples). The synthetic datasets could be used to evaluate the robustness and versatility of clustering algorithms.

The algorithm of text clustering dataset generator is shown in Appendix E. It should be illustrated that the embedding and LLM models are not restricted to the models evaluated in this paper. In principle, any compatible embedding model or LLM can be used in the generator. All user-defined parameters and their descriptions are listed below:

\begin{itemize}
    \item \textbf{Data}: The dataset for synthesis. The text dataset should be preprocessed to remove noisy or invalid characters.
    \item \textbf{Embedding model}: The encoder used to map text into latent representations, options include Instructor, E5, Qwen, Gemma, and SBERT.
    \item \textbf{LLM model}: The large language model new text samples generation, options include Qwen-8B and Llama-3.1-8B.
    \item \textbf{Number of new samples $M$}: The number of generated samples per selected label. By default, this matches the original sample size of the selected labels in the input dataset. For the imbalance type, the default range is random from $0.1\times$ to $n\times$ the original sample size of the selected labels, where $n$ is user-defined to control the imbalance ratio.
    \item \textbf{Number of labels}: The number of labels to be synthesized, default is all labels.
    \item \textbf{Selected label list}: An optional user-specified list of labels to synthesize.
    \item \textbf{Number of examples $K$}: The number of in-context examples used to construct each prompt.
    \item \textbf{Synthesis type (SynType)}: The synthesis operation, options include imbalance, compactness, and diversity.
    \item \textbf{Merge type}: Whether to merge generated samples with the original dataset or replace selected original samples.
\end{itemize}

\subsection{Encoder Experiments}

The text encoder is a critical component of the generator. We design some experiments to examine whether latent representations produced by different encoder-only models (e.g., E5 and Instructor) present a similar clustering structure. The results can validate the robustness of the generator to the choice of embedding methods.

\begin{table}[ht]
\centering
\caption{Embedding similarity between the Instructor and the other 4 encoders.}
\begin{tabular}{lccc}
\hline
 & Local & Pearson & MAD \\
\hline
E5    & 0.677 & 0.7538 & 0.0967 \\
Gemma & 0.578 & 0.6188 & 0.0817 \\
Qwen  & 0.591 & 0.7345 & 0.4812 \\
Sbert & 0.628 & 0.7329 & 0.5533 \\
\hline
\end{tabular}
\label{table2}
\end{table}

Table \ref{table2} compares the similarity between Instructor representations and those produced by four other encoder-only models in terms of local neighborhood overlap, Pearson correlation, and mean absolute difference as evaluation metrics (MAD). The details of these embedding models are listed in Appendix B.2.

The local neighborhood overlap is the most directly relevant metric to evaluate whether two embedding spaces have similar clustering structure. It measures the consistency of the local geometry between two embeddings by comparing their k-nearest neighbor sets (cosine distance, $k=10$). In this table, all four encoders achieve overlap rates above 0.5, indicating a generally similar local structure among models.

The Pearson correlation represents the global geometry between two embedding spaces. In Table \ref{table2}, all four encoders achieve a Pearson correlations above 0.5, demonstrating that their global geometry is reasonably consistent. We further report the mean absolute difference to quantify pairwise similarity between embedding spaces. The instructor has higher similarity with E5 and Gemma and lower similarity with Qwen and Sbert, suggesting notable differences in absolute distance scale and embedding-space density with the latter two models.

\begin{table}[ht]
\centering
\caption{Clustering Performance of synthesized Imbalance CLINC datasets from different embedding and LLM models.}
\begin{tabular}{c|lcc|ccc}
\hline
&Method & ARI & NMI & SI & DB & CH \\
\hline
\hline
\multirow{5}{*}{\rotatebox{90}{Qwen}}&Instructor   & 0.7158 & 0.9118 & 0.2632 & 2.0994 & 38.8126 \\
&E5        & 0.7002 & 0.9090 & 0.2591 & 2.0923 & 38.7458 \\
&Qwen         & 0.7287 & 0.9153 & 0.2635 & 2.0291 & 38.4655 \\
&Gemma        & 0.7177 & 0.9087 & 0.2633 & 2.1009 & 38.8549 \\
&Sbert        & 0.7425 & 0.9151 & 0.2682 & 2.1057 & 39.0415 \\
\hline
\multirow{5}{*}{\rotatebox{90}{Llama}}&Instructor  & 0.6747 & 0.8998 & 0.2514 & 2.1171 & 37.1655 \\
&E5          & 0.6997 & 0.9056 & 0.2491 & 2.1170 & 37.5616 \\
&Qwen        & 0.6678 & 0.8990 & 0.2352 & 2.1196 & 35.1765 \\
&Gemma       & 0.7021 & 0.9007 & 0.2392 & 2.1282 & 35.3796 \\
&Sbert      & 0.6745 & 0.8965 & 0.2441 & 2.1563 & 35.2711 \\
\hline
\end{tabular}
\label{table 3}
\end{table}

Overall, the selection of the encoder model has a limited impact on the quality of the generated datasets. Table \ref{table 3} further supports this conclusion. We encode the CLINC dataset using five encoder-only models (INSTRUCTOR, E5, Qwen, Gemma, and SBERT) and generate imbalance-type synthetic datasets by two LLMs (Qwen and Llama). The details of LLMs are presented in Appendix B.1. The generated datasets are evaluated with the $k$-means clustering method under the same parameters, in terms of ARI, Normalized Mutual Information (NMI), silhouette score (SI), Davies-Bouldin (DB), and Calinski-Harabasz (CH). The details of these metrics are shown in Appendix D.3. The results indicate that different encoders have a relatively small effect on the clusterability of the generated datasets. In contrast, the choice of LLM has a more remarkable influence. The datasets synthesized by Qwen consistently achieve a better clustering performance than those generated by Llama.

\begin{figure*}[!htbp]
    \centering

    \begin{subfigure}[t]{0.23\textwidth}
        \centering
        \includegraphics[width=\linewidth]{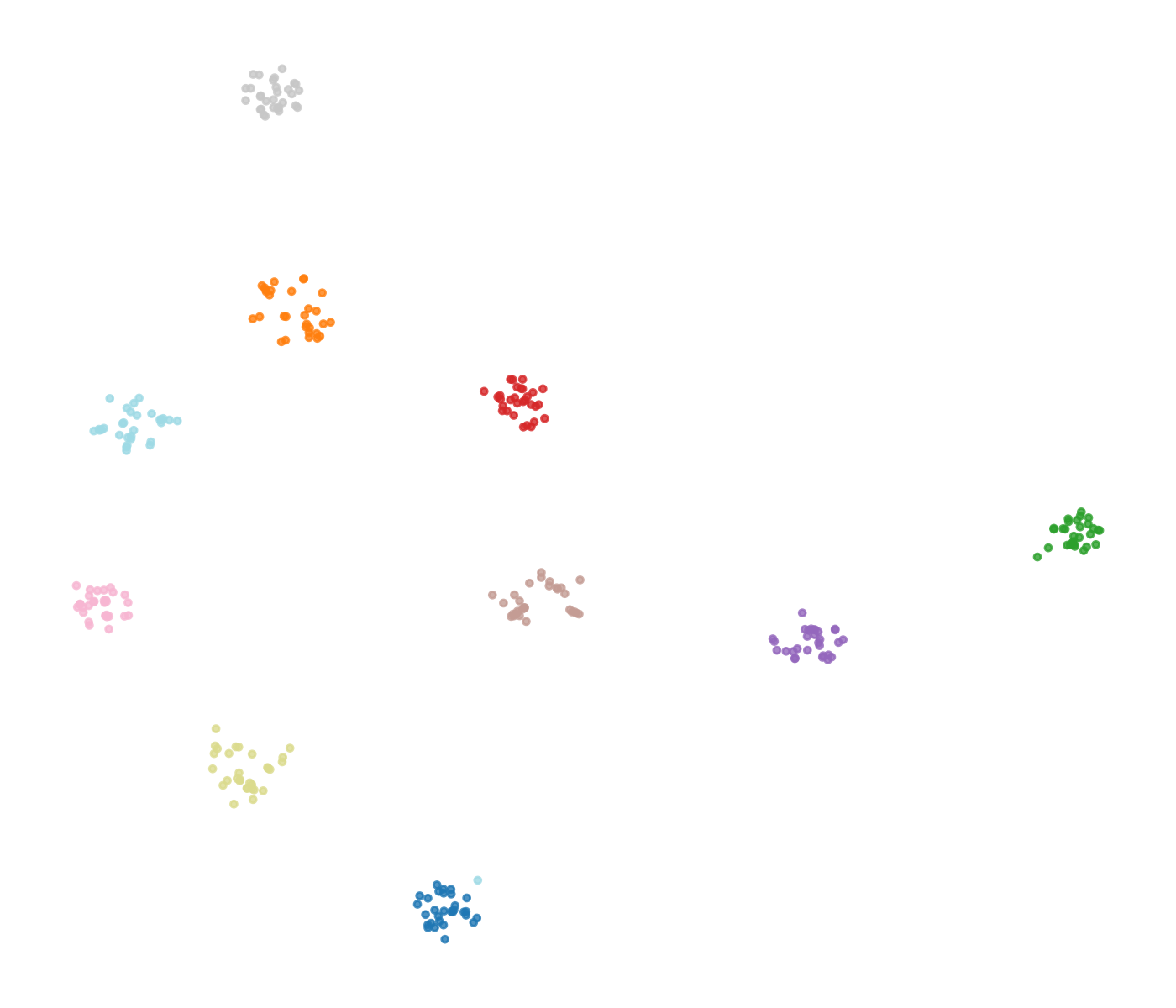}
        \caption{Unsupervised CLINC}
        \label{fig:balance_clinc}
    \end{subfigure}
    \hfill
    \begin{subfigure}[t]{0.23\textwidth}
        \centering
        \includegraphics[width=\linewidth]{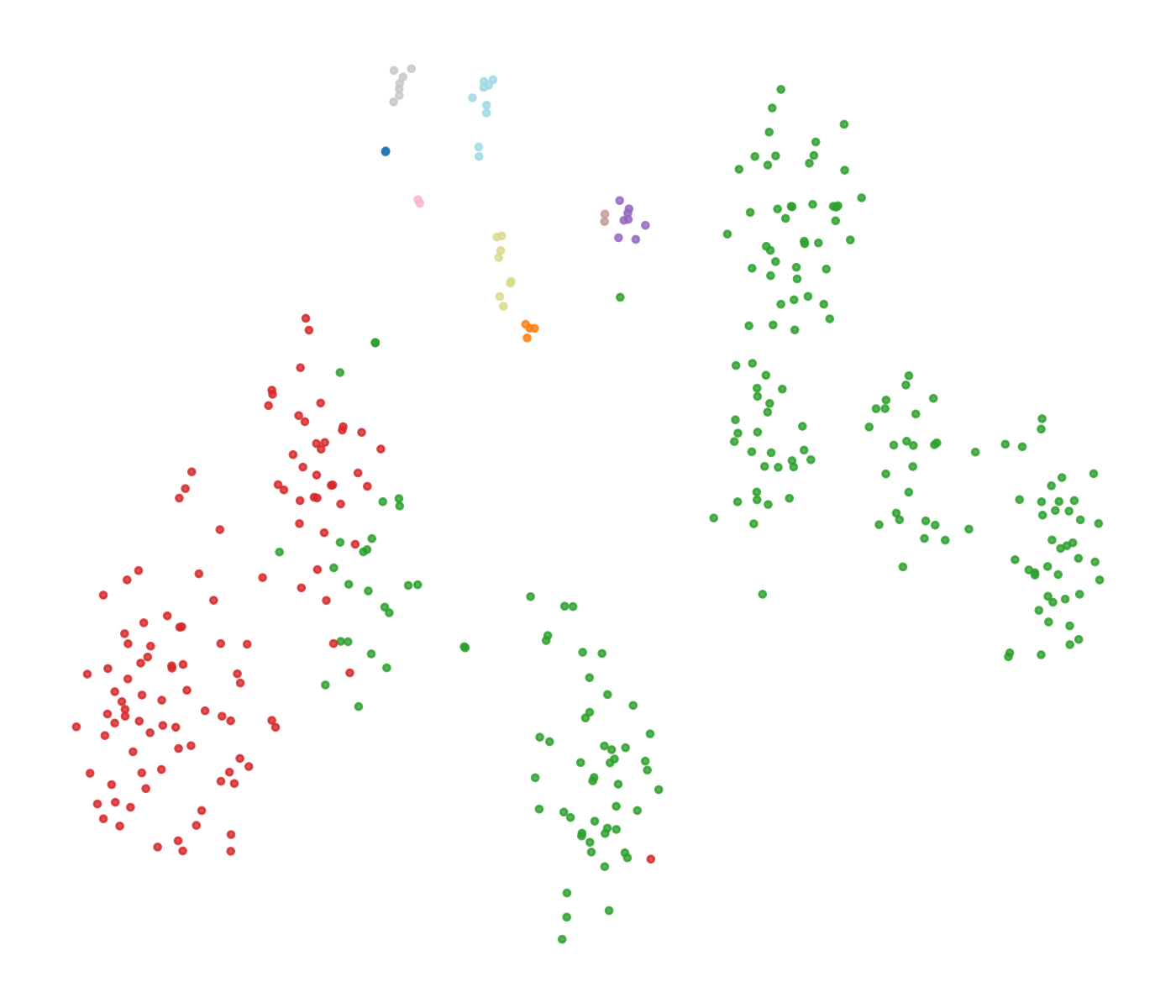}
        \caption{Mtop}
        \label{fig:Imbalance_mtop}
    \end{subfigure}\hfill
    \begin{subfigure}[t]{0.23\textwidth}
        \centering
        \includegraphics[width=\linewidth]{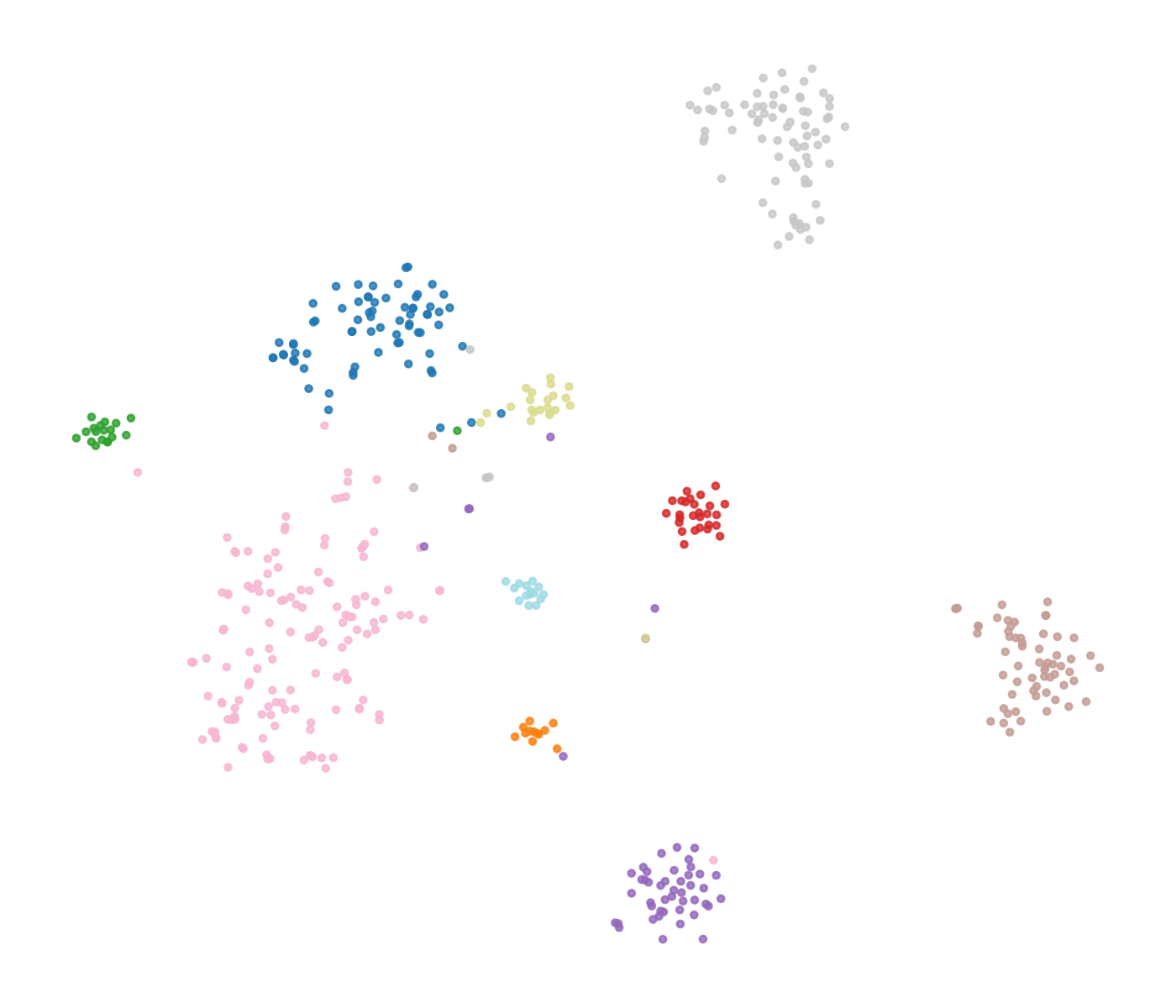}
        \caption{Massive}
        \label{fig:massive}
    \end{subfigure}\hfill
    \begin{subfigure}[t]{0.23\textwidth}
        \centering
        \includegraphics[width=\linewidth]{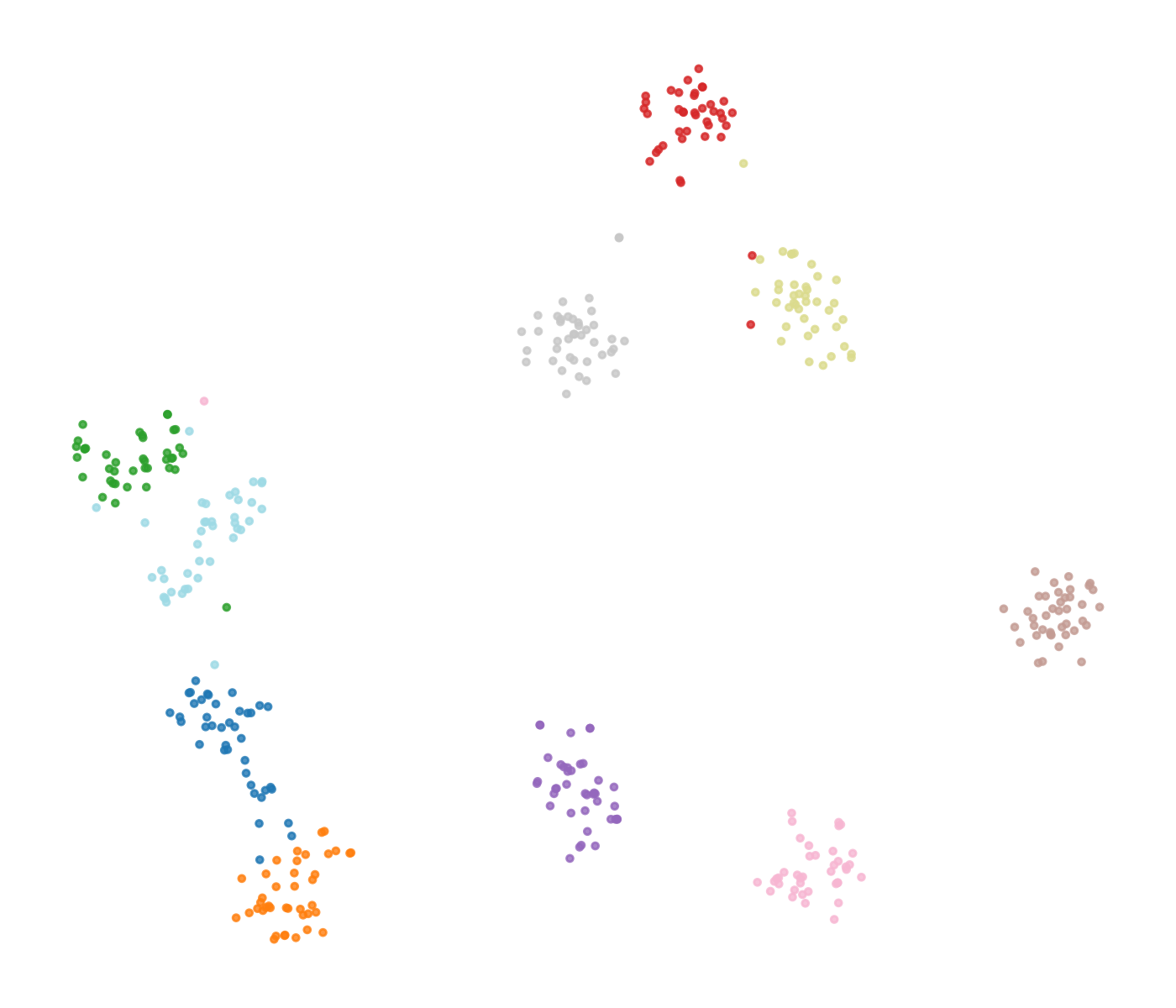}
        \caption{Banking77}
        \label{fig:banking77}
    \end{subfigure}

    \vspace{6pt} 

    \begin{subfigure}[t]{0.23\textwidth}
        \centering
        \includegraphics[width=\linewidth]{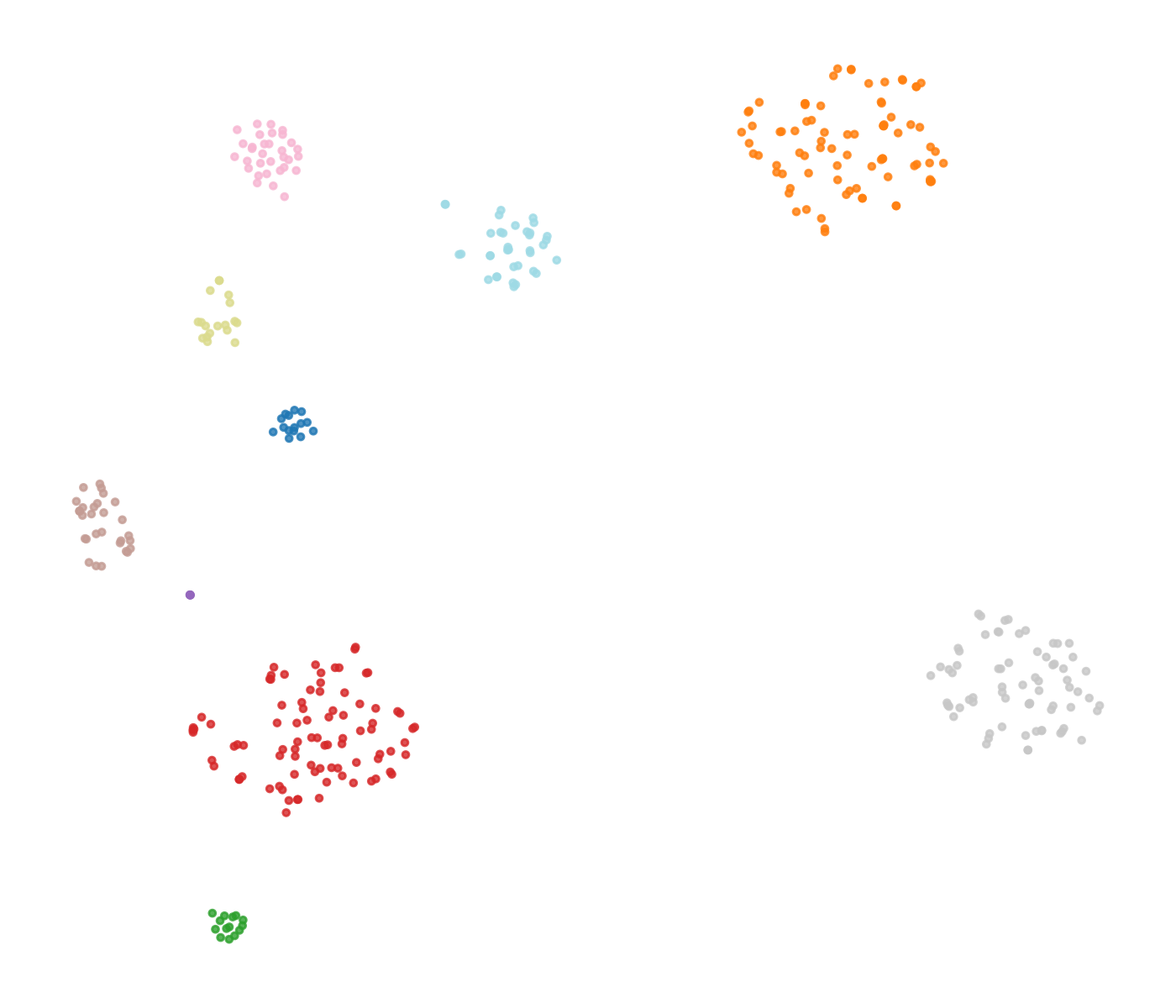}
        \caption{ Imbalance CLINC }
        \label{fig:Synthetic clinc}
    \end{subfigure}\hfill
    \begin{subfigure}[t]{0.23\textwidth}
        \centering
        \includegraphics[width=\linewidth]{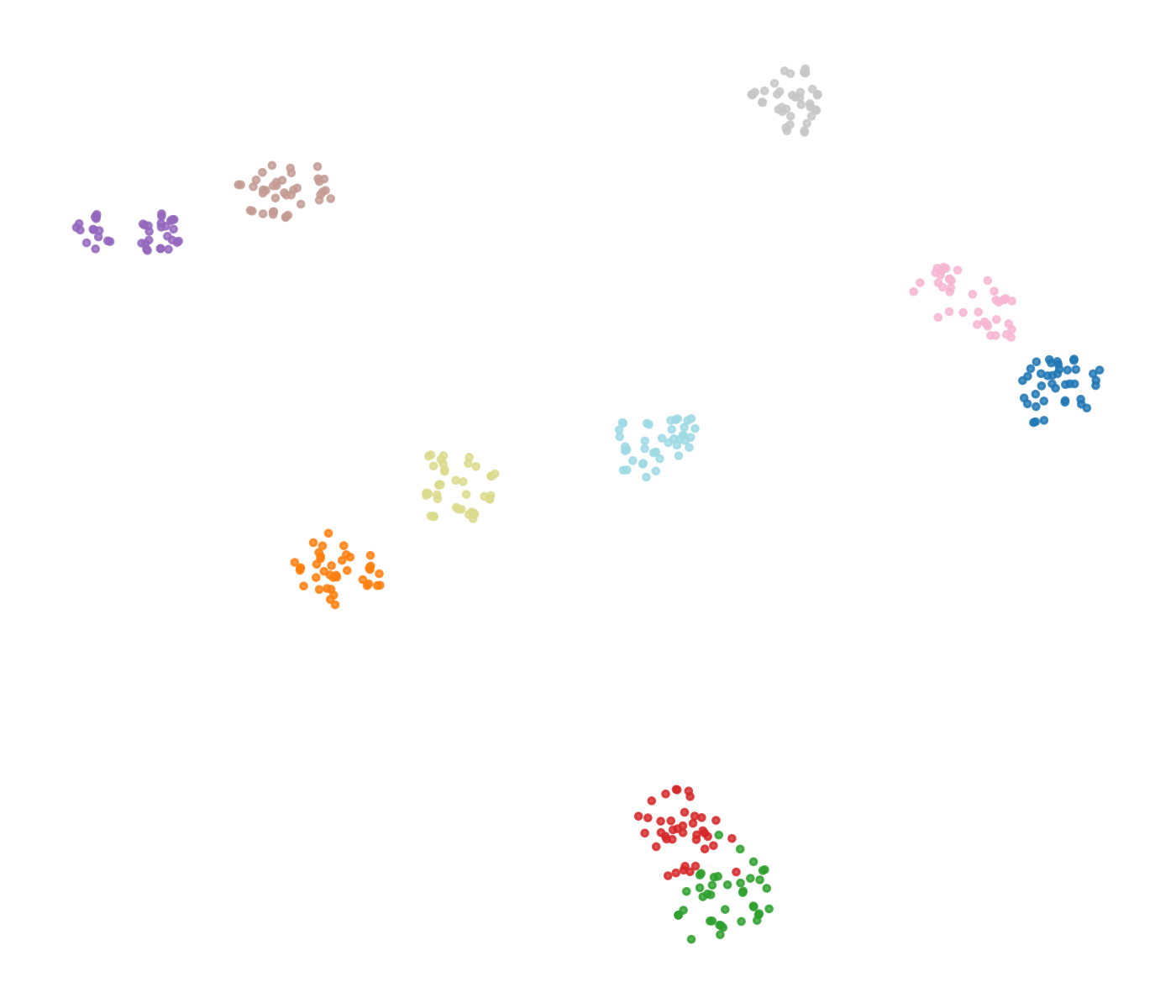}
        \caption{ Balance Mtop}
        \label{fig:Synthetic mtop}
    \end{subfigure}\hfill
    \begin{subfigure}[t]{0.23\textwidth}
        \centering
        \includegraphics[width=\linewidth]{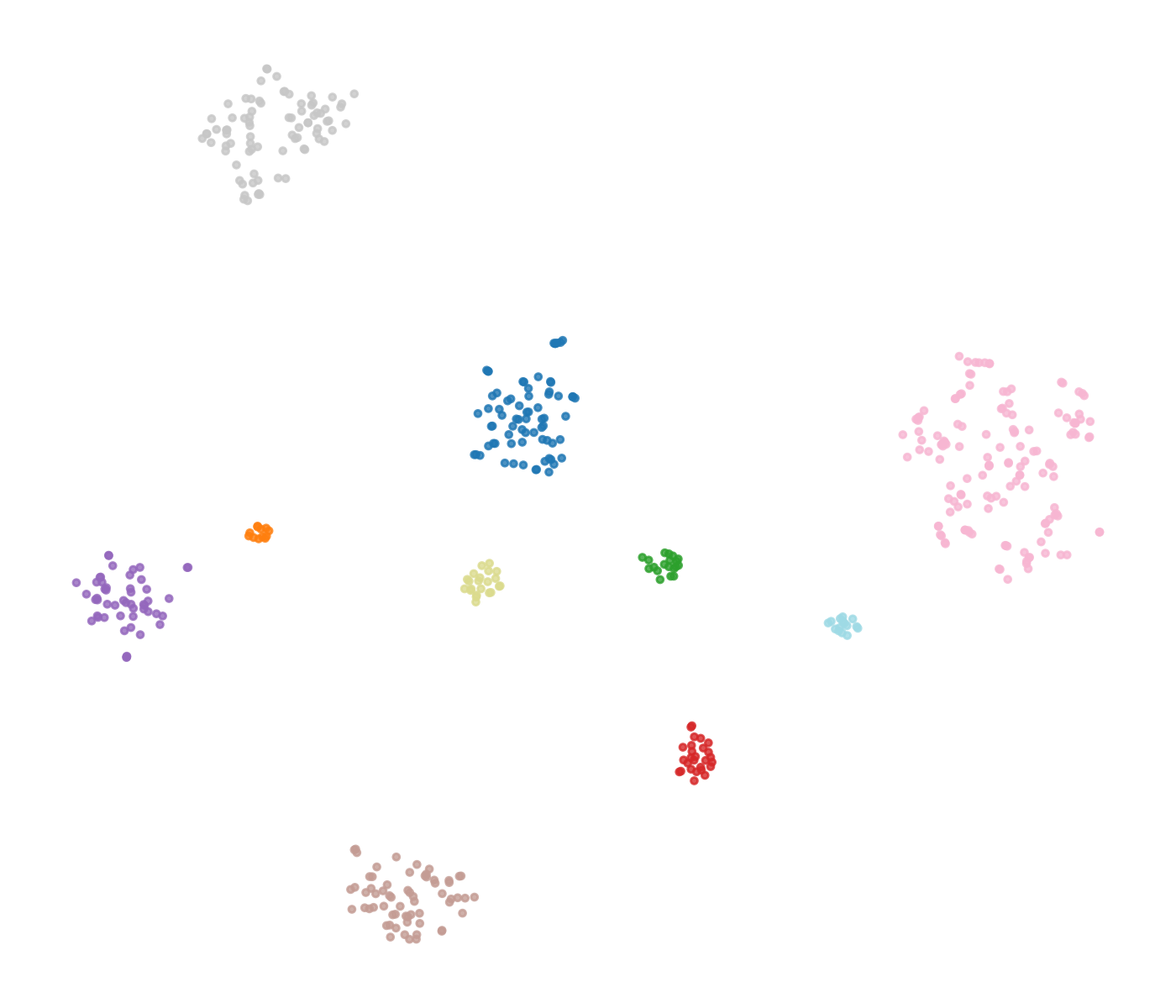}
        \caption{ Compact Massive}
        \label{fig:Synthetic massive}
    \end{subfigure}\hfill
    \begin{subfigure}[t]{0.23\textwidth}
        \centering
        \includegraphics[width=\linewidth]{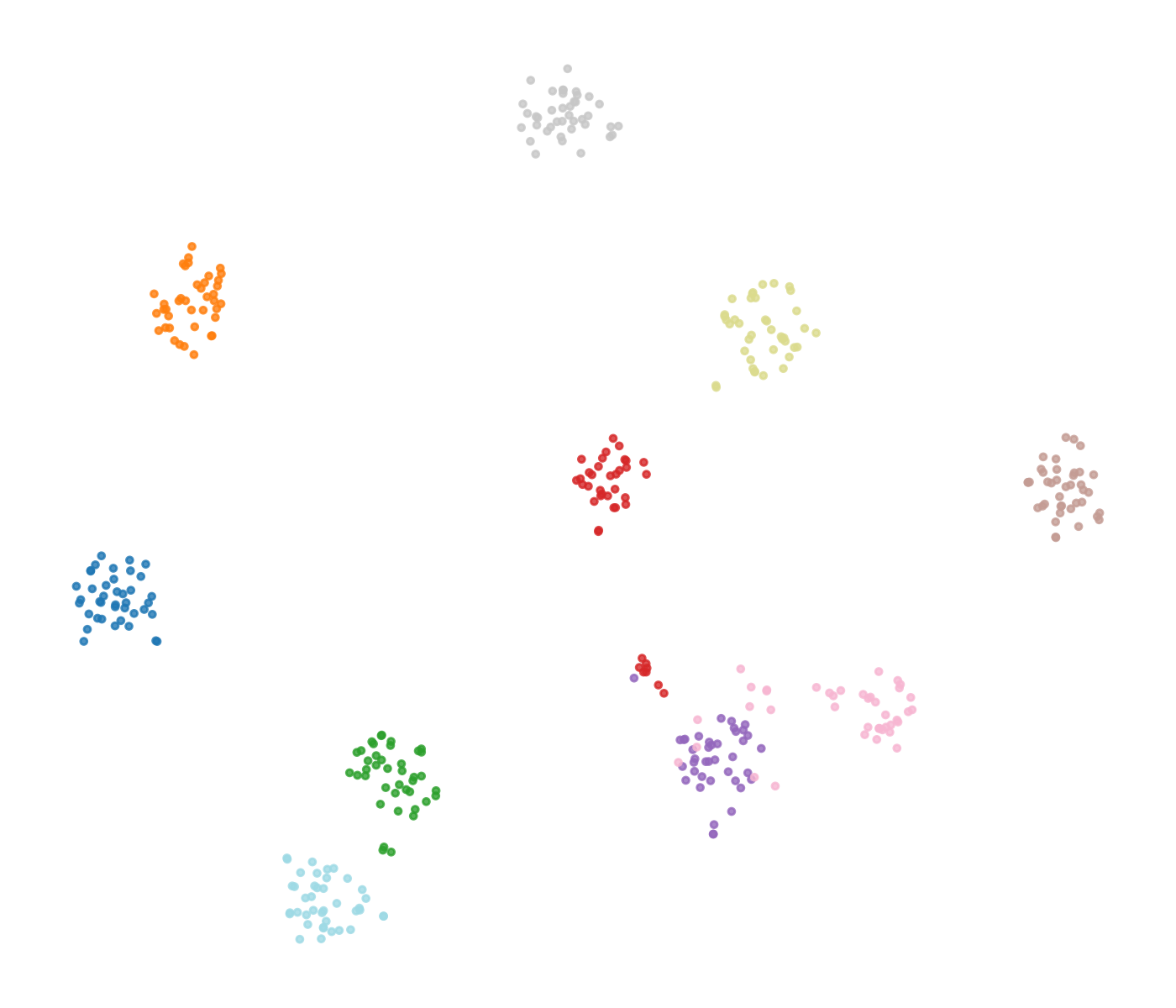}
        \caption{Diversity Banking77}
        \label{fig:Synthetic banking77}
    \end{subfigure}
    \caption{t-SNE visualization of the four extracted datasets (CLINC, Banking77, Massive, and Mtop) and their corresponding Synthetic datasets.}
    \label{figure3}
\end{figure*}

\subsection{Synthetic Dataset Comparison}

To clearly demonstrate the differences between synthetic data and original data, we consider four source datasets (CLINC, banking77, Massive, and Mtop). For each dataset, we extract 10 classes to facilitate visualization of cluster structure in the t-SNE plots. In addition, we remove the ground-truth labels from CLINC to simulate the generator pipeline under an unsupervised setting. All synthetic datasets are generated using the Instructor encoder and the Qwen LLM. The t-SNE visualizations are presented in Figure \ref{figure3}, the top four subfigures are original datasets, and the bottom four subfigures are synthetic datasets. Their corresponding $k$-means clustering metrics are reported in Table \ref{table4}.

For the CLINC dataset, the t-SNE visualizations indicate that the original (label-balanced) dataset has a tight cluster structure, whereas the imbalanced synthetic dataset shows a looser structure. To simulate an unsupervised setting, we measured only internal (label-free) metrics. The synthetic imbalanced dataset has lower SI and CH scores and a higher DB index. It supports the conclusion from t-SNE visualizations that the balanced dataset is more clusterable, with compact clusters and well-separated centroids.

\begin{table}[ht]
\caption{Clustering metrics for 4 original datasets and their corresponding synthetic datasets.}

\centering
\begin{tabular}{lccccc}
\toprule
Dataset & ARI & NMI & SI & DB & CH \\
\midrule
CLINC                &  &  & 0.5981 & 1.2587 & 62.7906 \\
Imba\_CLINC     & &  & 0.5558 & 1.4819 & 66.1869 \\
\hline
Mtop                 & 0.2367 & 0.5460 & 0.2253 & 2.0048 & 35.7698 \\
Balance\_Mtop        & 0.9779 & 0.9811 & 0.3982 & 1.7398 & 66.9892 \\
\hline
Massive              & 0.7492 & 0.8679 & 0.3413 & 1.9061 & 41.4540 \\
Comp\_Massive     & 0.8218 & 0.9338 & 0.4109 & 1.6931 & 61.4312 \\
\hline
Banking77            & 0.9348 & 0.9507 & 0.4156 & 1.5659 & 56.0721 \\
Div\_Banking77 & 0.8961 & 0.9476 & 0.4443 & 1.4272 & 63.3145 \\
\bottomrule
\label{table4}
\end{tabular}
\label{tab:clustering_metrics}

\end{table}

For the Mtop dataset, the original data are imbalanced, leading to low external scores (ARI and NMI). In contrast, the synthetic balanced dataset achieves comparatively higher external performance, and its internal metrics are also significantly improved compared with the original datasets, demonstrating a clearer cluster structure. This improvement is also observed in the t-SNE visualization. These results indicate that balanced synthetic datasets can improve the clusterability of the original data while preserving semantic consistency within each label, which is particularly beneficial for domains where datasets are inherently imbalanced.

For the Massive dataset, the synthetic compact dataset achieves slightly improved internal and external metrics compared with the original data. Although the original dataset is extremely imbalanced, the compactness-oriented synthesis helps to separate a broader cluster boundary and improve overall clusterability.

For the Banking77 dataset, the diversity-oriented synthetic data has lower external metrics but improved internal metrics. A reasonable explanation is that the introduced multi-semantic samples weaken label–cluster alignment. The boundary/overlap samples are ambiguous and difficult to assign to the correct cluster, which degrades external performance.

\begin{table}[h]
\centering
\caption{Clusterability criteria scores (C1--C8) for Massive and Compact Massive.}
\begin{tabular}{ccccc}
\toprule
     &Massive &  Compact Massive\\
\midrule
Reliability (C1) & 98 & 99 \\
Structure (C2)   & 87 & 92 \\
Stability (C3)   & 87 & 94 \\
Separation (C4)  & 52 & 65 \\
Agreement (C5)   & 57 & 67 \\
Coherence (C6)   & 72 & 75 \\
Imbalance (C7)   & 0 & 0 \\
Tendency (C8)    & 75 & 76 \\
\bottomrule
\label{table 5}
\end{tabular}
\end{table}

\begin{figure}[htbp]
  \centering
  \includegraphics[width=0.3\textwidth]{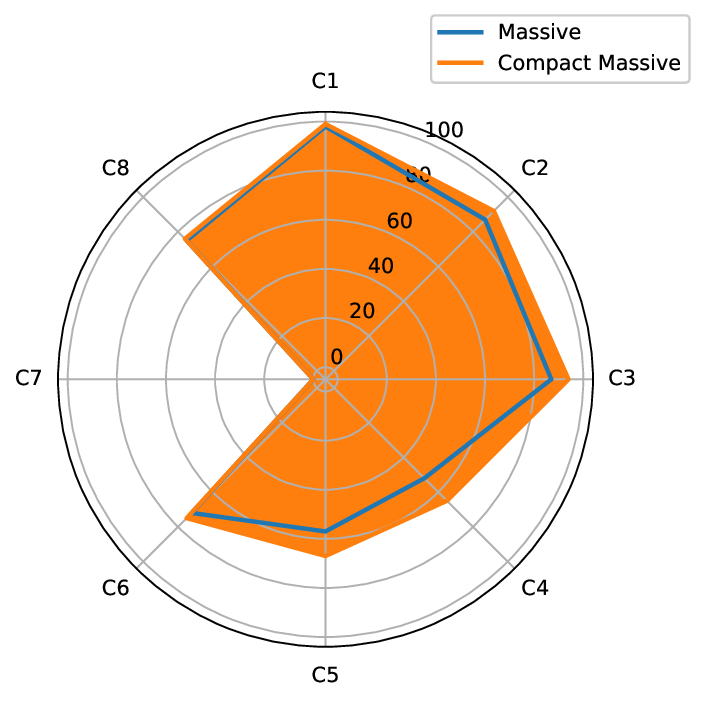}
    \caption{Benchmark radar of Massive and Compact Massive.}
    \label{figure 4}
\end{figure}

We next present an example to illustrate how the generator can improve dataset clusterability. Since the original Massive dataset scores weakly on the Separation criterion, we apply the compactness-oriented strategy to generate a synthetic dataset. Table \ref{table 5} shows that the compact synthetic dataset achieves higher scores on all criteria except Imbalance, since the generated dataset preserves the original sample size of each label. The radar chart in Figure \ref{figure 4} further supports this observation. The compact synthetic dataset presents consistently stronger scores, indicating improved clusterability compared to the original dataset.

\section{Evaluation Protocol}

Synthetic datasets can serve as a practical evaluation protocol for clustering methods. In some domains, data are rare or difficult to collect. Moreover, even when datasets are available, their cluster attributes are often homogeneous (i.e., they present limited variation in attributes such as imbalance, compactness, or boundary ambiguity). Clustering methods may perform well for these current limited properties, but generalize poorly to datasets encountered in the future. In this case, a text clustering dataset generator can synthesize diverse datasets with controlled attributes, helping to systematically evaluate the robustness and versatility of clustering methods for domain-specific applications.

\begin{table}[ht]
\centering
\caption{Clustering Performance of three clustering algorithms ($k$-means, HDBSCAN, and agglomerative) in four data types (Balance, Imbalance, Compact, and Diversity).}
\begin{tabular}{c|lcc|ccc}
\hline
&Method & ARI & NMI & SI & DB & CH \\
\hline
\hline
\multirow{5}{*}{\rotatebox{90}{$k$-means}}&Balance   & 0.6838 & 0.8644 & 0.1939 & 2.4592 & 42.5801 \\
&Imbalance        & 0.3279 & 0.7044 & 0.1411 & 2.7647 & 30.9483 \\
&Compact         & 0.6301 & 0.8666 & 0.2197 & 2.4757 & 49.3633 \\
&Diversity        & 0.5830 & 0.8557 & 0.1930 & 2.4282 & 49.3591 \\
\hline
\multirow{5}{*}{\rotatebox{90}{hdbscan}}&Balance  & 0.0330 & 0.5300 & -0.0831 & 2.0075 & 18.8518 \\
&Imbalance          & 0.0183 & 0.1929 & -0.1136 & 3.1082 & 15.2482 \\
&Compact        & 0.0886 & 0.6184 & -0.0174 & 1.9969 & 27.7275 \\
&Diversity       & 0.0713 & 0.5817 & -0.0494 & 2.0025 & 25.8006 \\
\hline
\multirow{5}{*}{\rotatebox{90}{agglo}}&Balance & 0.6699 & 0.8610 & 0.1916 & 2.5684 & 41.7330 \\
&Imbalance        & 0.3607 & 0.6971 & 0.1114 & 2.9131 & 29.3347 \\
&Compact         & 0.6539 & 0.8809 & 0.2269 & 2.3454 & 49.3773 \\
&Diversity        & 0.6140 & 0.8679 & 0.2027 & 2.3527 & 49.2139 \\
\hline
\end{tabular}
\label{table 6}
\end{table}

We present a specific example of three clustering algorithms ($k$-means, HDBSCAN, and agglomerative) on the Massive dataset. We use the generator to synthesize datasets with different clustering properties (balanced, compactness, and diversity). Because the original Massive dataset is extremely imbalanced, we adopt it unchanged as an additional test set. Table \ref{table 6} reports the performance of the three algorithms in terms of ARI, NMI, SI, DB, and CH. From Table \ref{table 6}, $k$-means and agglomerative clustering are obviously more robust in this domain. Both of them achieve strong performance on the synthetic balanced and compact datasets for both internal and external metrics. However, their performance degrades remarkably on the original imbalanced dataset and the diversity-oriented synthetic dataset, indicating limited versatility under more challenging clustering conditions (e.g., severe imbalance and boundary ambiguity). In contrast, HDBSCAN performs consistently poorly in all data properties, demonstrating that it is not well-suited for this domain under the current experimental configuration.

\section{Limitations}

The proposed framework provides a stable and effective foundation for generating and evaluating text clustering datasets. While our experiments demonstrate the practical value of the generator using specific open-source LLMs and embedding techniques, the rapid evolution of LLMs offers opportunities for continuous expansion. Our current validation focuses on a representative selection of models to ensure accessibility and reproducibility. Future research could extend this scope to include a wider range of architectures to further verify the adaptability of the framework across different model families.

The current application of TextClusterLab mainly targets short to medium-length texts such as user queries and intent classification data. These data types are common in many industrial applications and allow for efficient prompt construction within standard context windows. Processing significantly longer texts might require adjustments to the prompt engineering strategy to manage context limits effectively.

\section{Conclusions}

In this paper, we propose a novel clusterability benchmark for studying text datasets prior to evaluating clustering algorithms. We further introduce an LLM-driven generator that enables specific control over various clustering attributes. The produced synthetic datasets provide an evaluation protocol for testing the robustness and versatility of clustering methods under diverse clustering properties. Generally, these components constitute an integrated pipeline for systematic, reproducible text clustering research, a ``text clustering laboratory''.

\section{Acknowledgments}
This research is supported by Natural Sciences and Engineering Research Council (NSERC) of Canada and York Research Chairs (YRC) program. All the work was done when the first author was a Ph.D. student at the Information Retrieval and Knowledge Management Research Lab, York University, Canada.

\bibliographystyle{ACM-Reference-Format}
\bibliography{sample-base}

\clearpage
\appendix

\section{Prompt Settings}
To ensure reproducibility and facilitate the generation of high-quality synthetic datasets, we provide the system and user prompts used for the Qwen and Llama models below. These instructions are engineered to maximize the utility of the generated text for clustering benchmarks by enforcing three key constraints:

\begin{enumerate}
    \item \textbf{Format Compliance:} Outputs are restricted to valid JSON arrays to ensure robust automated parsing without post-processing errors.
    \item \textbf{Content Diversity:} We utilize a dynamic prompting strategy. In addition to instructing the model to avoid verbatim repetition of few-shot examples, we inject a dynamic ``avoid list'' into the user message containing recently generated utterances to prevent duplicates.
    \item \textbf{Structural Consistency:} Explicit constraints on word count and character length ensure the synthetic utterances mimic the short-text nature of real-world intent datasets.
\end{enumerate}

The specific prompts for Qwen and Llama are presented below.

\begin{promptbox}{Prompt 1: Input for Qwen Model}
\small
\textbf{[System Message]} \\
You generate synthetic intent-classification utterances.
Return ONLY the final answer.
Do NOT include analysis, thinking, or explanations.
Output must be a JSON array of strings (no markdown).
Return exactly $N$ strings.

HARD RULES for EACH string:
\begin{itemize}
    \item Do NOT copy any provided example verbatim.
    \item Keep same intent; vary wording/entities/length.
    \item Use normal spaces between words (no underscores ``\_'' and no ``\_'' markers).
    \item Each string must be a natural-language utterance (not character lists or repeated letters).
    \item Do NOT output strings that are mostly single-letter tokens or repeated characters.
    \item Each string should have at least 4 words.
    \item Do NOT output timestamps, IDs, counters, alphabet ranges, or percentage lists.
    \item Avoid bracketed character sequences like [A] [B] [C] or long symbol runs.
\end{itemize}

\vspace{0.5em}
\hrule
\vspace{0.5em}

\textbf{[User Message]} \\
Nonce: \{Random UUID\} \\
Intent label: \{Target Intent\} \\
\\
Examples (same intent):
\begin{itemize}
    \item \{Example 1\}
    \item \{Example 2\}
    \item ...
\end{itemize}

\textit{(Conditional Block: If avoid\_texts exists)} \\
Avoid repeating any of these exactly:
\begin{itemize}
    \item \{Recently Generated Text 1\}
    \item \{Recently Generated Text 2\}
    \item ...
\end{itemize}

Generate now as a JSON array of strings.
\end{promptbox}

\begin{promptbox}{Prompt 2: Input for Llama Model}
\small
\textbf{[System Message]} \\
You generate synthetic intent-classification utterances.
Return ONLY the final answer.
Output must be a JSON array of strings (no markdown).
Return exactly $N$ strings.

HARD RULES for EACH string:
\begin{itemize}
    \item One utterance only (single line). No newline characters.
    \item 6 to 20 words.
    \item Max 140 characters.
    \item Natural conversational English.
    \item Do NOT repeat any phrase of 4+ words.
    \item Do NOT include markup/tags/code (no \texttt{<...>}, no JSON objects).
    \item Do NOT include weird character runs (e.g., ``vvvvv'', ``a b c'').
    \item Do NOT copy any provided example verbatim.
    \item Use normal spaces between words (no underscores ``\_'' and no ``\_'' markers).
\end{itemize}
Return ONLY the JSON array. No extra text.

\vspace{0.5em}
\hrule
\vspace{0.5em}

\textbf{[User Message]} \\
Intent label: \{Target Intent\} \\
\\
Examples (same intent, for style only):
\begin{itemize}
    \item \{Example 1\}
    \item \{Example 2\}
    \item ...
\end{itemize}

\textit{(Conditional Block: If avoid\_texts exists)} \\
Avoid repeating any of these exactly:
\begin{itemize}
    \item \{Recently Generated Text 1\}
    \item \{Recently Generated Text 2\}
    \item ...
\end{itemize}

Now generate $N$ NEW utterances as a JSON array of strings.
\end{promptbox}

\section{Model Settings}
\subsection{Generation Models}
We employ two state-of-the-art open-weights LLMs to drive the \textit{TextClusterLab} generator. These models serve as the backbone for synthesizing intent-classification utterances based on the prompts defined in the previous section.

\paragraph{Qwen3 (Qwen/Qwen3-8b):} We utilize the Qwen family of models \cite{yang2025qwen3}, specifically the instruction-tuned versions. Qwen is selected for its strong performance in following complex formatting constraints (e.g., JSON output) and its multilingual capabilities. Trained on a massive corpus of approximately 36 trillion tokens, the Qwen series introduces advanced reasoning capabilities and significant improvements in coding and mathematics compared to its predecessors. Its architecture utilizes Grouped Query Attention and SwiGLU activation to maintain high inference efficiency while handling context lengths of up to 128k tokens.

\paragraph{Llama3.1 (Llama/Llama3.1-8b):} We utilize the Llama family of models \cite{grattafiori2024llama}, a widely adopted benchmark for open-source LLMs. Its robust natural language understanding ensures the synthetic utterances remain diverse and semantically coherent. The models are trained on over 15 trillion tokens of data and optimized via Supervised Fine-Tuning and Reinforcement Learning with Human Feedback to prioritize helpfulness and safety. With deep architectural optimizations like Grouped Query Attention and an 8B parameter scale, Llama excels at capturing nuance in conversational English, making it ideal for generating naturalistic intent utterances.

\subsection{Embedding Models}
To evaluate clustering performance across diverse latent space distributions, we select five distinct embedding models ranging from lightweight encoders to large-scale multilingual models. The specific configurations are listed below:

\paragraph{Instructor-Large (hku-nlp/instructor-large)} An instruction-finetuned embedding model with 335M parameters. It generates task-specific embeddings by prepending instructions to the input text \cite{su-etal-2023-one}. Initialized from the GTR-Large model, it is trained on a diverse mixture of 330 datasets across 70 distinct embedding tasks. This instruction-aware design allows the model to generate customized representations for specific downstream tasks, ensuring that embeddings capture the most relevant semantic features for clustering without requiring fine-tuning.

\paragraph{Multilingual-E5-Large (intfloat/multilingual-e5-large)} A 560M parameter model initialized from XLM-RoBERTa. It is trained on large-scale text pairs and supports more than 100 languages, providing a high-dimensional (1024d) and dense latent space \cite{wang2024multilingual}. The model utilizes a two-stage training pipeline consisting of weakly supervised contrastive pre-training on one billion text pairs followed by supervised fine-tuning. It requires specific prefixes to distinguish between symmetric and asymmetric tasks, enhancing its ability to separate semantically distinct clusters in multilingual environments.

\paragraph{Qwen3-Embedding (Qwen/Qwen3-Embedding-0.6B)} A 0.6B parameter model from the Qwen family. It supports a context length of up to 32k tokens and covers more than 100 languages, offering a balance between model size and representation capability in a 1024-dimensional space \cite{qwen3embedding}. This model integrates Matryoshka Representation Learning (MRL), allowing flexible output vector sizes without retraining, and supports instruction-based input to refine embedding quality by 1--5\% on downstream tasks. Despite its compact size, it leverages the dense training data of the Qwen3 foundation to achieve competitive performance on massive text embedding benchmarks (MTEB).

\paragraph{EmbeddingGemma (google/embeddinggemma-300m)} A 300M parameter model distilled from the Gemma family. Using MRL, it is designed for efficient and high-performance embedding generation with a compact memory footprint \cite{vera2025embeddinggemma0}. Built upon the Gemma 3 architecture, this model is specifically fine-tuned for semantic similarity and clustering tasks using a diverse mixture of synthetic and real-world data. It offers a 768-dimensional output that can be truncated for storage efficiency, making it highly suitable for resource-constrained clustering scenarios.

\paragraph{SBERT (sentence-transformers/all-MiniLM-L6-v2)} A highly efficient, lightweight model (approx. 22M parameters) that maps sentences to a 384-dimensional vector space. It serves as a baseline for performance in low-resource and low-latency scenarios \cite{reimers2019sentence0bert0}. The model is distilled from a larger BERT-based teacher using a 1-billion sentence pair dataset, optimizing for cosine similarity preservation. Its small size and fast inference speed make it a standard baseline for evaluating whether larger, more complex embedding models provide statistically significant gains in clustering quality.

\begin{table}[hb]
\centering
\caption{Hyperparameter settings for text generation models.}
\label{tab:param_settings}
\begin{tabular}{lcc}
\toprule
\textbf{Parameter} & \textbf{Qwen} & \textbf{Llama} \\
\midrule
Batch Size & 10 & 10 \\
Number of Shots & 10 & 8 \\
Avoid Last $k$ & 3 & 12 \\
Max Attempts & 60 & 60 \\
Temperature (Start) & 0.95 & 0.95 \\
Top-p (Start) & 0.85 & 0.8 \\
Presence Penalty & 0.6 & 0.25 \\
Frequency Penalty & 0.6 & 0.35 \\
Max Tokens & 2400 & 2400 \\
Seed & 42 & 42 \\
\bottomrule
\end{tabular}
\end{table}

\begin{table*}[ht]
  \centering
  \caption{Comparison of Original Samples vs. LLM-Synthesized Samples for the ``Bill Balance'' Intent.}
  \label{tab:data_comparison_full}
  
  \renewcommand{\arraystretch}{1.25}
  
  \resizebox{\textwidth}{!}{
  \begin{tabular}{|l|l|}
    \hline
    \textbf{Original Data} & \textbf{Synthetic Generation} \\
    \hline
    
    \textsf{what do i owe to jcp} & what is the current balance of my electric and gas bill \\
    \textsf{what is the amount due on my visa} & can you show me the total amount due for my water and internet \\
    \textsf{how much do i owe visa} & i would like to check the remaining balance on my phone and cable services \\
    \textsf{how much do you think my rent and electric bills are} & what is the outstanding amount for my property tax and insurance \\
    \textsf{what would you say my gas and phone bills are at} & how much do i still owe for my trash and recycling charges \\
    \textsf{i need to know the price of my car payment and insurance please} & could you tell me the current balance of my parking and maintenance fees \\
    \textsf{what price is my credit card and my water bill at} & i need to see the remaining amount for my security deposit and utility charges \\
    \textsf{how much is my water and sewer} & what is the total due for my internet and mobile phone services this month \\
    \textsf{how much do i need to pay for my electricity and water bills} & how much is left to pay for my car insurance and loan installments \\
    \textsf{what is the cost of my rent and water bills} & can you provide the current balance of my health insurance premium and service charges \\
    \textsf{how much are my rent and cable} & what is the current balance of my credit card \\
    \textsf{how much does my water and electricity cost} & can you show me the total amount due for my utilities this month \\
    \textsf{what do i owe this month on all my bills} & i would like to check the remaining balance on my phone bill \\
    \textsf{i think all my bills are paid, but can you double check} & how much is the outstanding amount for my internet service \\
    \textsf{i think i owe about 100 left on my bills, am i forgetting anything} & what is the current status of my loan payment balance \\
    \textsf{what are my bills this month} & could you tell me the total amount owed for my electricity and gas bills \\
    \textsf{what does my cable bill look like} & i need to check the balance of my recent car insurance payment \\
    \textsf{has my electricity bill increased this month how much} & how much is the remaining amount on my previous month's water bill \\
    \textsf{how expensive is my internet} & what is the current total of all my pending bills \\
    \textsf{how much will my monthly bill be} & can you provide the balance information for my latest phone and internet charges \\
    \textsf{what are my total bills this month} & what is the current balance of my electric bill \\
    \textsf{what will i be paying for utilities this month} & can you show me the total amount due for my phone service \\
    \textsf{how much is my telephone bill this month} & i would like to check the outstanding amount on my credit card statement \\
    \textsf{how much is due for my water bill} & how much is the sum of my recent gas and internet charges \\
    \textsf{how much do i have to pay this month} & what is the total of my current water and electricity bills \\
    \textsf{tell me my water bill please} & could you tell me the amount i owe for my car insurance and registration \\
    \textsf{is the phone bill the same as last month} & i need to find out the total payment required for my property tax and maintenance \\
    \textsf{whens my insurance due and how much does it cost} & how much is the combined cost of my mobile and landline services this month \\
    \textsf{how much is the car bill this month} & what is the total amount due for my recent medical and dental expenses \\
    \textsf{what is the electricity bill} & can you provide the current balance of my loan and interest payments \\
    
    \hline
  \end{tabular}
  }
  \label{table 8}
\end{table*}

\section{Parameter Settings}
To ensure reproducibility and control the quality of synthetic data, we configure the generation hyperparameters. Note that parameters controlling stochasticity and penalties are set as initial values, which the generation loop may dynamically adjust if the model fails to produce valid output after multiple attempts. The specific definitions are as follows:

\begin{itemize}
    \item \textbf{Batch Size:} The number of prompt requests processed in parallel to maximize generation throughput.
    \item \textbf{N Shots:} The number of few-shot examples included in the context window to guide the model's output style and format.
    \item \textbf{Avoid Last $k$:} A constraint that prevents the model from generating an utterance identical to any of the immediately preceding $k$ samples.
    \item \textbf{Max Attempts:} The maximum number of retry iterations allowed if the model produces malformed, duplicate, or non-compliant output.
    \item \textbf{Temperature (Start):} The initial coefficient for controlling randomness in token selection; higher values yield more diverse outputs, while lower values are more deterministic.
    \item \textbf{Top-p (Start):} The initial threshold for nucleus sampling, restricting the candidate pool to the smallest set of tokens with a cumulative probability exceeding $p$.
    \item \textbf{Presence Penalty:} The initial penalty applied to tokens that have already appeared anywhere in the output, encouraging the model to introduce new topics.
    \item \textbf{Frequency Penalty:} The initial penalty scaled by how many times a token has appeared, discouraging the model from verbatim repetition or looping of phrases.
    \item \textbf{Max Tokens:} The hard limit on the number of tokens in the generated response to enforce conciseness.
    \item \textbf{Seed:} The fixed random state used to initialize the model for reproducible results.
\end{itemize}

We configure the generation parameters for Qwen and Llama to optimize for diversity while adhering to formatting constraints. Qwen is configured with a larger batch size and context window (max tokens) due to its robustness in following complex instructions, while Llama uses a smaller batch size and tighter token limits to ensure high-quality, concise outputs. The specific hyperparameter settings are detailed in Table \ref{tab:param_settings}.

\begin{table}[h]
\centering
\caption{Statistics of the datasets used in experiments.}
\label{tab:dataset_stats}
\begin{tabular}{lcccc}
\toprule
\textbf{Dataset} & \textbf{Classes} & \textbf{Type} & \textbf{Challenge} \\
\midrule
CLINC   & 150 & Balanced & Multi-domain separation \\
Banking77  & 77  & Balanced & Fine-grained similarity \\
MASSIVE    & 59  & Imbalanced & Real-world skew \\
MTOP       & 104 & Imbalanced & Extreme skew \\
\bottomrule
\end{tabular}
\end{table}

\section{Dataset Settings}

\subsection{Dataset Samples Comparison}
Table \ref{tab:data_comparison_full} presents examples from the original dataset and the synthetic dataset. All examples belong to the ``Bill Balance'' label. The synthetic text is not a rewrite of the original text in the same row.

\subsection{Experiment Datasets}
To evaluate the proposed framework, we select four datasets that represent different clustering challenges, including fine-grained semantic overlap and extreme class imbalance. The key statistics for these datasets are summarized in Table \ref{tab:dataset_stats}.

\paragraph{CLINC} The CLINC dataset \cite{larson-etal-2019-evaluation} covers 150 diverse intent classes across 10 domains. We utilize the in-scope queries to form a balanced dataset. It serves as a standard benchmark for evaluating clustering performance on categories that are relatively distinct and well-separated.

\paragraph{Banking77} The Banking77 dataset \cite{casanueva-etal-2020-efficient} contains 77 fine-grained intents within the banking domain. Unlike CLINC, this dataset has high semantic overlap between classes (e.g., distinct intents for ``card arrival'' vs. ``card delivery''). It tests the ability of the model to distinguish subtle semantic differences in a dense latent space.

\paragraph{MASSIVE} The MASSIVE dataset \cite{fitzgerald-etal-2023-massive} is a multilingual NLU benchmark. We utilize the English subset for our experiments. A key characteristic of MASSIVE is its real-world distribution; it is highly imbalanced, with significant variance in sample counts per intent. This allows us to evaluate the robustness of the generator and clustering algorithms under label skew.

\paragraph{MTOP} The MTOP dataset \cite{li-etal-2021-mtop} is derived from task-oriented semantic parsing queries. Similar to MASSIVE, it presents a highly imbalanced structure. As shown in our clusterability benchmark (Table \ref{table 1}), it contains classes ranging from 1 sample to nearly 500 samples. This extreme imbalance provides a rigorous test for handling density variations in clustering.

\subsection{Dataset Metrics}

We employ five standard metrics to assess the generated datasets and clustering algorithm performance. These include external metrics that require ground-truth labels and internal metrics that evaluate the latent space structure without external references.

\textbf{The Adjusted Rand Index (ARI)} acts as an external measure of similarity between ground-truth labels and clustering assignments. It calculates the frequency of sample pairs assigned to the same or different clusters in both partitions. This metric corrects the standard Rand Index for chance and ensures that random assignment yields a score near zero. A score of 1 indicates a perfect match while lower values suggest poor agreement.

\textbf{Normalized Mutual Information (NMI)} quantifies shared information between ground-truth labels and clustering results. It measures the reduction in entropy of class labels given cluster assignments. The score ranges from 0 to 1 where 1 indicates perfect correlation. This metric is independent of absolute label values and allows comparison across different datasets.

\textbf{The Silhouette Score (SI)} evaluates how well an object fits within its cluster. It compares the mean distance between a sample and all other points in the same cluster to the mean distance to points in the nearest neighboring cluster. Values range from -1 to 1 where a high score suggests the object matches its own cluster well. A value near 0 indicates overlapping clusters while negative values imply incorrect assignment.

\textbf{The Davies-Bouldin Index (DB)} calculates the average similarity between each cluster and its most similar counterpart based on the ratio of within-cluster scatter to between-cluster separation. A lower index indicates better performance with compact and distinct clusters. This score reflects a model with low intra-cluster variance and high inter-cluster distance.

\textbf{The Calinski-Harabasz Index (CH)} evaluates the ratio of the sum of between-cluster dispersion to within-cluster dispersion for all clusters. A higher score implies that clusters are dense and well-separated. This metric is computationally efficient and helps identify convex clusters. We use it to verify if synthetic datasets maintain defined boundaries in the high-dimensional latent space.

\section{Algorithm}
The algorithm of text clustering dataset generator is shown in Algorithm \ref{A1}.

\begin{algorithm}[h]
\normalsize
\caption{Text Clustering Dataset Generator}
\begin{algorithmic}[1]
\REQUIRE Text data ${D} = \{ x_i \}_{i=1}^{n}$, new sample size $M$, select label set $L$, example size $K$, ModType, and MergeType.
\ENSURE Synthetic dataset $\tilde{D}$

\STATE Initialize the pseudo-label if ${D}$ is unsupervised.
\STATE Encoding ${D}$ into $E = f(D)$.

\FOR{$c \in L$}
    \STATE Extract samples $S_c$ from the select label set $c$.
    \FOR{$x_i \in S_c$}
        \STATE Compute sample weight $w_i$.
        \IF {ModType = imbalance}
            \STATE $w_i = 1$
        \ENDIF
        \IF {ModType = compact}
            \STATE get $w_i$ by eq \ref{eq12}
        \ENDIF
        \IF {ModType = diversity}
            \STATE get $w_i$ by eq \ref{eq13}
        \ENDIF
        \STATE obtain normalized sampling distribution $p_i$ by eq \ref{eq14}.
    \ENDFOR
    \STATE Weighted Random Selection $K$ examples.
    \STATE Construct Prompt $P$ by $K$ examples.
    \STATE generate $M$ new samples $\tilde{S_c}$ by LLM.
\ENDFOR
\STATE Merge or Duplicate $\tilde{S_c}$ with ${D}$ to get Synthetic dataset $\tilde{D}$.
\end{algorithmic}
\label{A1}
\end{algorithm}

\end{document}